# First-principles studies of spin-phonon coupling in $Cr_2Ge_2Te_6$ monolayer

B. H. Zhang,[1] Y. S. Hou,[2] Z. Wang,[1] R. Q. Wu[2*]

[1]*State Key Laboratory of Surface Physics, Key Laboratory of Computational Physical Sciences, and Department of Physics, Fudan University, Shanghai 200433, China*
[2]*Department of Physics and Astronomy, University of California, Irvine, California 92697, USA*

**Abstract**

We perform systematic first-principles calculations to investigate the spin-phonon coupling (SPC) of $Cr_2Ge_2Te_6$ (CGT) monolayer (ML). It is found that the $E_g$ phonon mode at 211.8 cm$^{-1}$ may have a SPC as large as 3.19 cm$^{-1}$, as it directly alters the superexchange interaction along the Cr-Te-Cr pathway. Furthermore, the strength of SPC of the CGT ML can be further enhanced by an in-plane compressive strain. These results provide useful insights for the understanding of SPC in novel two-dimensional magnetic semiconductors and may guide the design of spintronic and spin Seebeck materials and devices.

Email: wur@uci.edu

## I. INTRODUCTION

Two-dimensional (2D) van der Waals (vdW) monolayers (MLs) have attracted considerable research interest in the last decade because of their unique physical property, adaptability, and high integrability with other materials for the design of novel electronic and spintronic devices [1-6]. One of the challenges is to generate stable magnetization in these materials at a reasonably high temperature so that they can be used to filter spin currents or to magnetize topological surface states, to name a few possibilities [7,8]. While magnetism can be induced in many nonmagnetic 2D materials such as graphene, hex-BN, and $MoS_2$ by introducing dopants, defects and edges, [9,10], several new ferromagnetic (FM) vdW monolayers such as recently synthesized $Cr_2Ge_2Te_6$ (CGT) [1] and $CrI_3$ [11] appear to be more attractive for fundamental studies as well as for applications [12], which has inspired a new wave of research interest in exploring and manipulating 2D magnetic materials [13,14]. Due to the presence of sizeable magnetic anisotropy energy, the FM ordering of these monolayers may sustain at finite temperature, unlike what was predicted by the Mermin-Wigner theorem in the isotropic Heisenberg model [9]. Many experiments later have demonstrated that CGT and related 2D vdW magnetic MLs have many other remarkable merits [1,15-17], such as a large Kerr rotation angle, a giant modulation of the channel resistance, and a high thermoelectric performance. Integrating CGT with a topological insulator $Bi_2Te_3$ was found to greatly improve the electrical transport properties and enlarge the anomalous Hall conductivity [18].

To control and utilize the ferromagnetism of CGT and other 2D vdW magnetic MLs, a comprehensive study of their spin-phonon coupling (SPC) is crucial, because SPC plays an important role in magnetic fluctuation, magnonic dissipation, as well as in establishing the long-range FM order in these materials [3,19-28]. As far as we know, no quantitative theoretical study has been done for the SPC in 2D magnetic MLs, and only a handful experimental measurements of the SPC in CGT ML (e.g., using high resolution micro Raman scattering [3,20,24,25,27] and inelastic neutron scattering [21]) have been attempted [2]. This calls for more systematic and quantitative investigations for SPC in these new systems, so as to establish rules for guiding the 2D magnetic materials design.

In this work, we used the first-principles calculations to systematically investigate the SPC in CGT ML. Our results show that the SPC of the CGT ML varies with different vibrational modes. Especially, the $E_g$ mode at 211.8 cm$^{-1}$ shows the largest SPC, up to 3.19 cm$^{-1}$, which can be understood as a result of the strong interplay between the super exchange interaction through the Cr-Te-Cr path and the lattice deformation. Furthermore, we demonstrated that the SPC can be significantly enhanced by an in-plane compressive strain. Our work thus provides a deep insight regarding the SPC in 2D vdW magnetic monolayers and should be useful for their exploitation in various devices.

## II. METHODS

Our first-principles calculations were performed using the projector augmented wave (PAW) method as implemented in the Vienna *ab initio* simulation package (VASP). The exchange-correlation interactions were described by the generalized-gradient approximation (GGA) with the Perdew-Burke-Ernzerhof (PBE) functional [29]. An energy cutoff of 450 eV was used for the plane wave basis expansion. All atomic positions were fully relaxed until the force acting on each atom became smaller than 1×10$^{-5}$ eV/Å. We used the experimental lattice constant in the lateral plane for the CGT ML, *a*=*b*=6.83 Å, and adopted a slab model with a vacuum space of 17 Å to avoid the interaction between periodic images. The Brillouin zone was sampled with a 12×12×1 k-point mesh in the primitive cell. A 6×6×1 mesh was used for the 2×2×1 supercell. In calculating the phonon spectrum, we used a 4×4×1 supercell and a 3×3×1 k-point mesh. The strong correction effect among the 3*d* electrons of Cr was described by the LDA + U method, with an onsite coulomb and exchange parameters U=1.5 eV and J=0.5 eV respectively [1,15,30].

## III. RESULTS

### A. Magnetic Properties of 2D ML CGT

The CGT ML is a 2D FM semiconductor, and can be mechanically exfoliated from the bulk CGT [31]. Fig. 1(a)-1(b) show the crystal structure of the CGT ML. There are two Cr, two Ge and six Te atoms in the unit cell (u.c.). The magnetic Cr ions form a honeycomb lattice and locate at the center of a slightly distorted octahedron formed by six Te atoms. The total energy of the FM state (-47.155 eV/u.c.) is much lower than those of three other magnetic states, namely Néel antiferromagnetic (AFM) (Néel-AFM, -47.090 eV/u.c.), stripy AFM (stripy-AFM, -47.118 eV/u.c.) and zigzag AFM (zigzag-AFM, -47.140 eV/u.c.) [Fig. 1(c–f)]. This clearly shows that the FM order is the ground state of the CGT ML, consistent with experimental observations [32].

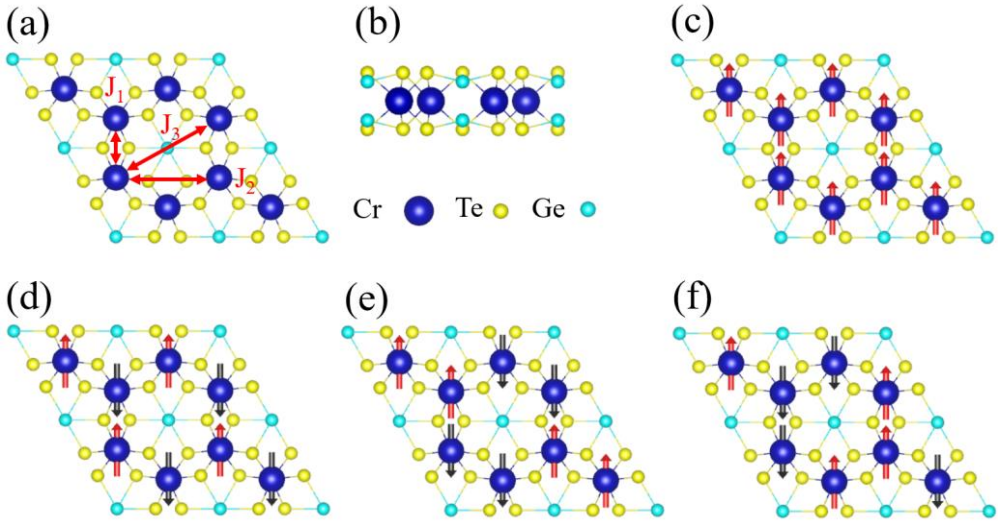

FIG. 1 (a) Top view and (b) side view of ML CGT. The considered exchange paths are shown by the red double arrows in (a). The possible magnetic configurations of ML CGT are (c) FM, (d) Néel-AFM, (e) stripy-AFM, and (f) zigzag-AFM orders.

Our calculations of the magnetic anisotropy energy (MAE) show that CGT ML with the experimental lattice constant has an out-of-plane easy axis. The magnetic anisotropy energy of CGT ML is composed of magnetocrystalline anisotropy energy (C-MAE) and magnetic dipolar anisotropy energy (D-MAE) which always prefers an in-plane magnetization. As depicted in Fig. 2(a), C-MAE turns from out-of-plane to the in-plane as the lattice constant increases. Consequently, the magnetic easy axis changes from out-of-plane to in-plane at the compressive strain of 0.41%. Besides, because C-AME depends strongly on the lattice constant but D-MAE has a weak dependence on the lattice constant, so it is feasible to enhance the out-of-plane magnetic anisotropy of the CGT ML through a compressive strain. Actually, there are other ways to enhance the out-of-plane magnetic anisotropy of the CGT ML, such as constructing heterostructures and using absorbates, as proposed in the literature [33,34,35,36].

To investigate the spin excitations, we first determine the exchange interactions between the magnetic ions Cr, utilizing a spin Hamiltonian as

$$H = -J_1 \sum_{\langle ij \rangle} \vec{S}_i \cdot \vec{S}_j - J_2 \sum_{\langle\langle ij \rangle\rangle} \vec{S}_i \cdot \vec{S}_j - J_3 \sum_{\langle\langle\langle ij \rangle\rangle\rangle} \vec{S}_i \cdot \vec{S}_j \qquad (1).$$

Here, $J_1$, $J_2$, and $J_3$ are the exchange interactions between the nearest neighbors (NN), second nearest neighbors (NNN), and third nearest neighbors (NNNN) (Fig. 1(a)). Note that positive values of $J_i (i=1,2,3)$ mean FM interactions and negative ones mean AFM interactions. Based on Eq. (1), the energy of the spin orders shown in Fig. 1c-1f are given as follows:

$$E_{\text{FM/Neel}} = E_0 \pm 12 J_1 |\vec{S}|^2 + 24 J_2 |\vec{S}|^2 \pm 12 J_3 |\vec{S}|^2 \qquad (2),$$

$$E_{\text{stripy/zigzag}} = E_0 \mp 4 J_1 |\vec{S}|^2 - 8 J_2 |\vec{S}|^2 \pm 12 J_3 |\vec{S}|^2 \qquad (3).$$

In Eq. (2) and (3), spin S for each Cr atom is 3/2. Our calculations show that $J_1$=4.92 meV, $J_2$=-0.31 meV, and $J_3$=0.01 meV, which agrees well with previous studies [15,37]. The large FM $J_1$ results from the competition between the direct exchange between Cr–Cr sites and the superexchange mediated through the Te ions, and strongly depends on the Cr-Cr distance. It is conceivable that $J_1$ changes when the CGT lattice is either stretched or compressed. This is confirmed in Fig. 2(a) which shows that $J_1$, $J_2$ and $J_3$ increase rapidly with lattice expansion in a range from 6.69 Å to 6.97 Å. Especially, the slope of the black curve in Fig. 2(a) indicates that the coefficient of the variation of $J_1$ with lattice expansion is as large as 21.13 meV/Å. Furthermore, it is obvious that the FM state becomes more stable when the lattice is stretched, because the FM $J_1$ increases but the AFM $J_2$ and $J_3$ decrease.

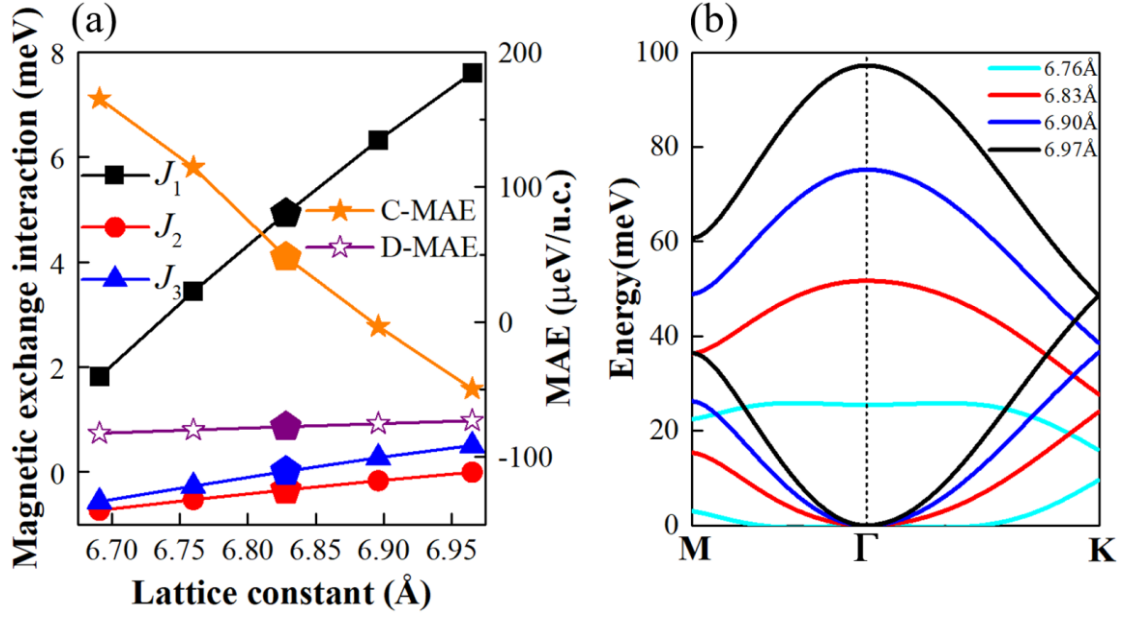

FIG. 2 *The dependences of (a) the magnetic exchange interactions, magnetic anisotropy energies and (b) spin waves on the lattice constant. The pentagonal symbols in (a) and the red lines in (b) give the magnetic anisotropy energy and the spin wave of ML CGT with the experimental lattice constant.*

To study the spin wave of CGT ML, we recast Eq. (1) in terms of the spin ladder operators $\hat{S}^{\pm} = \hat{S}^x \pm i\hat{S}^y$ and $\hat{S}^z$. For the its low-energy excitations, we obtain the spin wave spectrum as Eq. (4), by introducing the Holstein-Primakoff transformation [38] and Fourier transform (details are given in Part I in the Supplemental Materials).

$$\hbar\omega = 2S\sum_{n=1}^{3} Z_n J_n \pm 2S |\sum_{n=1}^{3} J_n \sum_{\delta_n} e^{i\vec{k}\cdot\vec{\delta}_n})| \qquad (4).$$

Here, $Z_1 = 3$, $Z_2 = 6$, and $Z_3 = 3$ are the first, second and third coordination numbers. The acoustic and optical modes of spin waves are obtained when the plus and minus sign are taken, respectively. As shown in Fig. 2(b), the energy of spin waves increases as the lattice constant increases. Note that the spin waves with a lattice constant a=6.69 Å is not included in Fig. 2(b) since the CGT ML becomes AFM state in this case. Again, the stiffness of spin waves and magnetic ground state of the CGT ML can be tuned by applying a strain, and strong spin-phonon coupling is expected [37].

## B. Phonon Properties of 2D ML CGT

The calculated phonon band dispersion of the 2D CGT ML is shown in Fig. 3(a). There is no negative phonon branch, indicating that this system is dynamically stable. Furthermore, 2D CGT ML is thermally stable, demonstrated by the ab initio dynamics simulations (see Part II in the Supplemental Materials). A remarkable feature of the phonon dispersion is the abundance of flat bands, which can also be appreciated from the numerous sharp peaks in phonon density of states (PDOS) as shown in Fig. 3(b). The PDOSs of the low-frequency modes (<150 cm$^{-1}$) are mainly dominated by the Te atoms due to their larger mass, whereas the mid-frequency modes (180~240 cm$^{-1}$) mainly result from vibrations of Cr atoms. The high-frequency modes (>260 cm$^{-1}$) mainly involve motions of Ge atoms. The flat bands in Fig. 3(a) and the weak coupling between different atoms in Fig. 3(b) clearly suggest that most vibration modes are rather local. As the ratio of Cr-motion is low in the low-energy acoustic modes, we focus on the optical modes around 200 cm$^{-1}$ for studies of SPC because there is a large ratio of Cr-motion. In Table 1, we give the irreducible symmetry representations of vibration modes of the CGT ML with the point group S$_6$, which can be expressed as $\Gamma=A_{1u}+3A_{2u}+3A_{1g}+2A_{2g}+5E_g+4E_u$. The infrared-active vibrational modes have the A$_{1u}$, 3A$_{2u}$, and 4E$_u$ symmetries, whereas the Raman-active modes have the 3A$_{1g}$, 2A$_{2g}$, and 5E$_g$ symmetries.

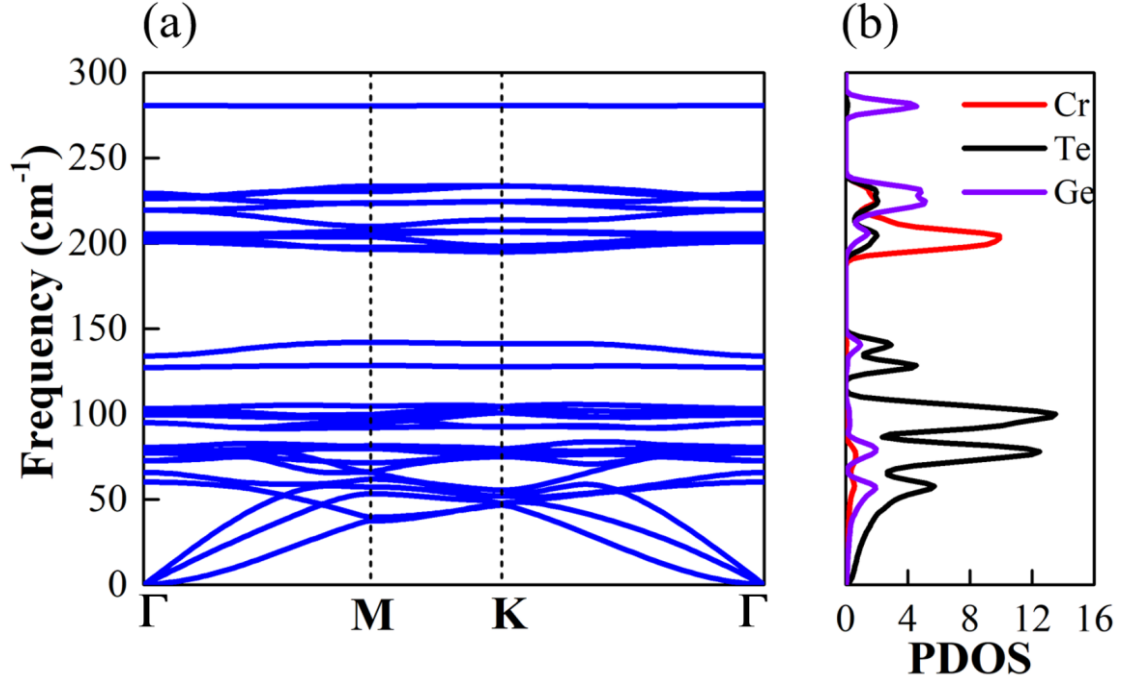

FIG. 3 (a) Phonon spectrum and (b) PDOSs of the CGT ML. The red, black and violet lines correspond to the PDOSs of Cr, Te and Ge, respectively.

Table1. Frequencies of the phonon vibrational modes for CGT ML at the $\Gamma$ point.

| Mode | Frequency (cm$^{-1}$) | Symmetry | Mode | Frequency (cm$^{-1}$) | symmetry |
|---|---|---|---|---|---|
| 1 | 59.38 | $A_{2u}$ | 15 | 130.06 | $A_{1g}$ |
| 2 | 70.24 | $A_{2g}$ | 16 | 140.17 | $A_{2u}$ |
| 3,4 | 73.11 | $E_g$ | 17,18 | 205.96 | $E_u$ |
| 5,6 | 78.12 | $E_u$ | 19 | 211.06 | $A_{2g}$ |
| 7,8 | 80.84 | $E_g$ | 20,21 | 211.81 | $E_g$ |
| 9,10 | 97.99 | $E_u$ | 22,23 | 226.01 | $E_g$ |
| 11 | 102.65 | $A_{1u}$ | 24,25 | 232.35 | $E_u$ |
| 12 | 104.75 | $A_{1g}$ | 26 | 238.38 | $A_{2u}$ |
| 13,14 | 106.15 | $E_g$ | 27 | 282.03 | $A_{1g}$ |

Here we focus on the Raman-active modes since they are measured in experiments [3]. The eigenvectors of these modes with the $3A_{1g}$, $2A_{2g}$, and $5E_g$ symmetries are schematically shown in Fig. 4. For the degenerate modes, we only show one branch of them. The modes in Fig. 4(a), 4(c) and 4(d) are mainly the vibrations of Te atoms around Cr atoms. The $E_g$ mode in Fig. 4(e) and $A_{1g}$ mode in Fig. 4(f) are the strong vibrations of Te atoms and tiny vibrations of Ge atoms. Note that the $E_g$ mode at 211.81 cm$^{-1}$ in Fig. 4(h) mainly involves anti-phase motions of the two Cr atoms. The $A_{2g}$ mode at 211.06 cm$^{-1}$ in Fig. 4(g) is out-of-plane vibrations of two Cr atoms, and in this case the Cr atoms move out-of-phase within the unit cell. Moreover, the $E_g$ and $A_{1g}$ modes shown in Fig. 4(i) and 4(j) consist of in-plane and out-of-plane vibrations of Ge layers, respectively.

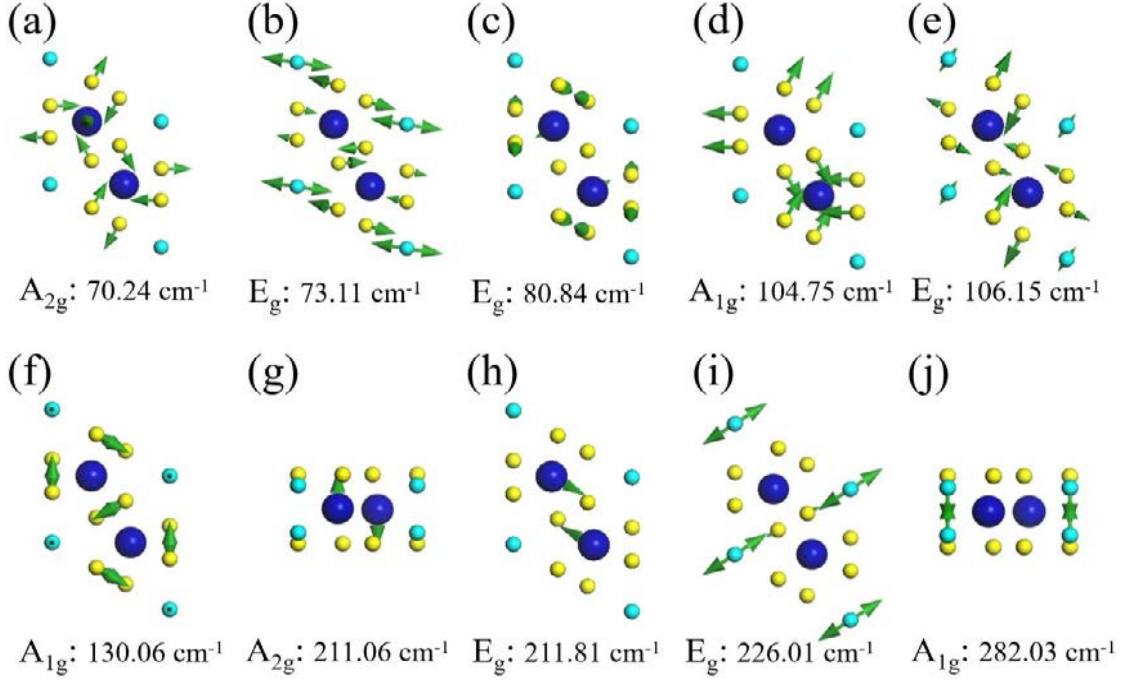

FIG.4 Schematic representations of the eigenvectors for all Raman active distinct phonon vibrational modes at the $\Gamma$ point. Cr, Ge and Te atoms are represented by the blue, cyan and yellow balls. The arrows show the directions and amplitudes of the atomic displacements in each mode.

### C. SPC of MLCGT

Now we investigate the SPC in the CGT ML with the frozen magnon approach proposed in Ref. [23] and Ref. [26]. The total energy $E$ can be expressed as:

$$E = E_0 - \sum_{ij} J_{ij} \vec{S}_i \cdot \vec{S}_j + \frac{1}{2} \sum_{\eta\eta'} (\tilde{C}_{\eta,\eta'} - \sum_{ij} \frac{\partial^2 J_{ij}}{\partial u_\eta \partial u_{\eta'}} \langle \vec{S}_i \cdot \vec{S}_j \rangle) u_\eta u_{\eta'}, \quad \eta \equiv \{n, \alpha\} \qquad (5).$$

In Eq. (5), the first and second terms are the paramagnetic (PM) energy and the

9 / 14

exchange interaction energy without vibrations and the third term is the energy due to the presence of phonon, respectively. $\eta$ represents the $\alpha$ direction ($\alpha=x, y, z$) of the $n$-th atom. For $J_{ij}$, we only consider $J_1$ since it plays a major role and it depends on the positions of N magnetic and nonmagnetic ions. $\tilde{C}_{\eta,\eta'}$ is the force constant for a particular magnetic ordering. $\partial^2 J_{ij}/\partial u_\eta \partial u_{\eta'}$ is the second derivative of the exchange interaction $J_1$ with respect to atomic displacements. The sums for $i$ and $j$ run over the magnetic atoms. In the CGT ML, we obtain $\tilde{C}^{FM}=C-6J''S^2$ and $\tilde{C}^{AFM}=C+6J''S^2$, where $\tilde{C}^{FM}$ and $\tilde{C}^{AFM}$ are the force constant matrices obtained by the first-principles calculations in the cases of the FM and AFM states. $C$ is the force constant for the PM phase. Therefore, the coupling parameter $\lambda_i$ for each mode $\psi_i$ can be easily calculated according to the expression:

$$\lambda_i = \langle \psi_i | \frac{-3J''}{S^2 \omega_i \sqrt{m_i m_j}} | \psi_i \rangle \qquad (6).$$

Here, $\omega_i$ represent the vibrational frequencies of PM states, and $m_i$ and $m_j$ are the mass of atoms.

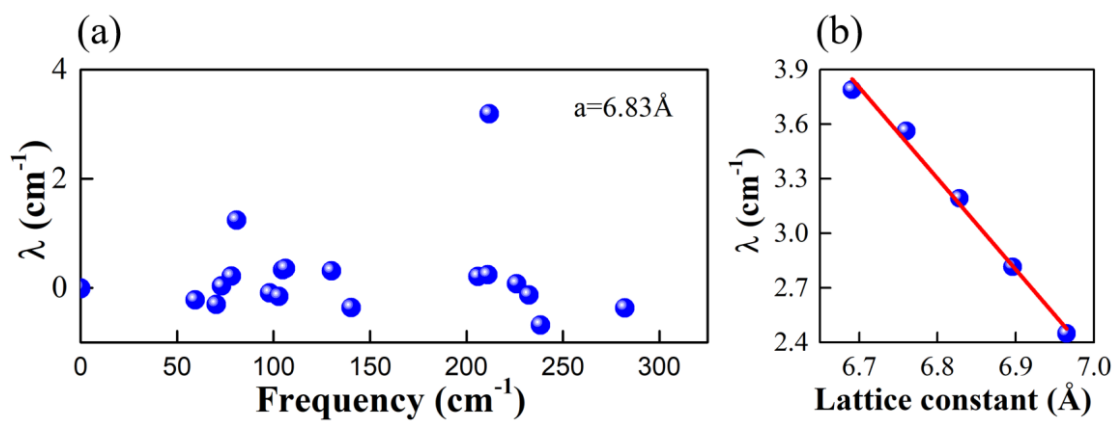

FIG.5 (a) The dispersion of spin–phonon coupling in the case of the experimental lattice constant a=6.83 Å. (b) The dependence of the SPC parameter at the 211.81 cm$^{-1}$ mode on the lattice constant.

Fig. 5(a) show the SPC coefficients of the CGT ML with the experimental lattice constant in different modes (The SPC of other lattice constants is shown in Part III in the Supplemental Materials). It is clear that the strength of SPC depends on the involvement of Cr-motion in the vibration model. For example, the $E_g$ mode at 211.81 cm$^{-1}$ that has a large Cr-component in Fig. 3(b) shows the largest SPC, about 3.19 cm$^{-1}$. The other sizable SPC is about 1.24 cm$^{-1}$, appearing in the $E_g$ mode at 80.84 cm$^{-1}$, which contains tiny vibrations of Cr atoms. As the $A_{2g}$ mode at 211.06 cm$^{-1}$ is out-of-plane vibrations of two Cr atoms, the SPC in this mode is small, about 0.24 cm$^{-1}$. Most of the remaining Raman-active modes have weak SPC coefficients since they mainly involve the vibrations of either Te or Ge atoms. Importantly, the SPC coefficient at the 211.81 cm$^{-1}$ mode change rapidly with the lattice strain, as shown in Fig. 5(b). Nonetheless, its dependence of the lattice constant is opposite to that of $J_1$, as SPC of the CGT ML enhances in the compression side. This is understandable since the CGT ML is about to have a phase transition from FM to AFM as the lattice size is close to 6.69 Å.

Now let us dig deeper into the SPC with respect to the atomic motions in different vibrational modes. It turns out that the $E_g$ mode at 211.81 cm$^{-1}$ in Fig. 4(h) contribute a more evident SPC since its vibrational modes is mainly involving Cr atoms which carry most of the magnetic moments. One observes that this vibrational mode mainly affects the bond length and angle of Cr-Te-Cr. Since the FM ground state of the ML CGT is mostly determined by the superexchange interaction along the Cr-Te-Cr path, this vibration changes $J_1$ greatly and contributes a strong SPC. Consequently, when the bond length and angle of Cr-Te-Cr is changed through the in-plane strain, the SPC also changes, as demonstrated above.

As a comparison, we also studied the SCP of on CrI$_3$ ML (see Part IV in Supplemental Materials). The calculated exchange interactions of CrI$_3$ ML ($J_1$=2.67 meV, $J_2$=0.64 meV and $J_3$= -0.12meV in experimental lattice constant [39]) are smaller than that of the CGT ML. We find that the SPC differs significantly in the two systems. The $E_g$ mode which mainly involves in-plane out-of-phase vibrations of two Cr atoms in the $CrI_3$ ML has the largest SPC (about 0.73 cm$^{-1}$). Obviously, one may qualitatively perceive the strength of SPC from the degree of deformation of superexchange path and the change of distance between magnetic atoms.

## IV. CONCLUSIONS

By means of the first-principles calculations, we systematically investigated the magnetic exchange interaction, spin waves, phonons and SPC of the 2D magnetic CGT ML. The magnetic exchange interaction $J$ and spin wave stiffness quickly increase as lattice constant expands. This provides a way to tune the magnetic properties of the CGT and other 2D magnetic MLs by strain and suggests the strong SPC in these materials. Frozen magnon calculations indicate that the $E_g$ phonon mode at 211.81 cm$^{-1}$ has the largest SPC, about 3.19 cm$^{-1}$, because it directly changes the bond length and angle of Cr-Te-Cr path and a large ratio of Cr-motion is included in this phonon mode. This study gives useful insights for the control of magnetic and phononic properties of 2D magnetic materials. The tunable SPC may find use in many applications such spin filtering, spin Seebeck, and spin wave control, to name a few.

**Acknowledgements**

The work at the University of California at Irvine was supported by the US DOE-BES under Grant DE-FG02-05ER46237. BZ and ZW also acknowledge support by National Basic Research Program of China under Grant No. 2015CB921400.